\documentclass[a4paper,10pt]{article}
\usepackage[utf8]{inputenc}
\usepackage{default}
\usepackage{hyperref}
\hypersetup{colorlinks=false, urlcolor=black}
\usepackage[letterpaper, margin=1.0in]{geometry}
\usepackage{graphicx}
\usepackage{epsfig}

\usepackage[round]{natbib}
\usepackage{enumerate}
\usepackage{microtype} 
\usepackage{subcaption}
\usepackage{tabularx}
\usepackage{booktabs}
\usepackage{csquotes}
\RequirePackage{fix-cm}

\usepackage{mathtext}
\usepackage{caption}
\usepackage[T1]{fontenc}    
\usepackage[english]{babel} 
\usepackage{amsmath, amsfonts, amssymb}	
\usepackage{latexsym}

\usepackage{appendix}
\usepackage{parskip}
\usepackage{lmodern}
\usepackage{authblk}  

\newcommand{\grad}{\hspace{-2mm}$\phantom{a}^{\circ}$}

\usepackage{xcolor}
\hypersetup{
    colorlinks,
    linkcolor={blue!80!black},
    citecolor={blue!80!black},
    urlcolor={blue!80!black}
}

\title{Evolution of Extreme Madden-Julian Oscillation Events and their Impacts on South America.}
\author[1]{M\'{o}nica Minjares}
\author[1,2,3]{\'{A}lvaro Corral}
\author[4]{Marcelo Barreiro}

\affil[1]{Departament de Física, Facultat de Ciències,Universitat Autònoma de Barcelona, E-08193 Barcelona, Spain}
\affil[2]{Centre de Recerca Matemática, Edifici C, Campus Bellaterra, E-08193 Barcelona, Spain}
\affil[3]{Complexity Science Hub Vienna,  Josefstädter Strasse 39, 1080 Vienna, Austria}
\affil[4]{Departamento de Ciencias de la Atm\'{o}sfera y Física de los Océanos, Facultad de Ciencias, Universidad de la Rep\'{u}blica, Igua 4225, 11400 Montevideo, Uruguay}

\date{}                     
\begin{document}

\maketitle


\noindent \textbf{Abstract}: This study examines the evolution of extreme Madden-Julian Oscillation (MJO) events and their impacts on South America during the austral summer. Furthermore, the study explores how the different ENSO phases modulate extreme MJO events, and how the combined effects impact South American climate. Extreme MJO events are defined as those exceeding a specific threshold based on the events distribution, distinguishing them from weak events. Our analysis shows that extreme MJO events most frequently initiate in phases 2-3 throughout the year, with similar distributions across phases 8-1, 6-7, and 4-5. This distribution is also characteristic of winter, while in summer, initiation is more balanced between phases 2-3 and 8-1. In contrast, weak events predominantly start in phases 4-5 year-round, followed by phases 2-3, with phases 8-1 and 6-7 occurring at similar frequencies. Seasonally, weak event initiation prevails in phases 4-5 during summer, while in winter, it is evenly distributed between phases 8-1 and 4-5. Additionally, during La Niña, extreme events tend to last longer than during El Niño, a pattern not observed in weak events.
A composite analysis of outgoing longwave radiation (OLR), eddy streamfunction, and velocity potential was conducted, with particular focus on the initiation phases 2-3 and 6-7 to determine which phases result in the most significant impacts and how the associated anomalies evolve. The findings show that enhanced (suppressed) convection centers in the equatorial region during extreme events are more intense and exhibit a southeastern displacement compared to those during weak events. These extreme MJO events influence the South American rainfall Dipole (SAD), a key feature of regional climate variability and results show that extreme MJO events induce more intense rainfall anomalies of larger spatial extent compared to weak events. This is because extreme events are characterized by dominant tropical-extratropical teleconnections, whereas weak events primarily feature tropical-tropical teleconnections. Also, as extreme events last longer, events that start in phases 2-3 influence the evolution of the SAD during the following phases, suggesting increased predictability of rainfall over South America.


\vspace{1 cm}

\section{Introduction}

MJO is the leading mode of intraseasonal variability in the tropics, having large influence on the tropical circulation and precipitation within the band of 30-60 days and is the most important source of global predictability on subseasonal time scales. MJO moves eastward around the globe, it is characterized by 2 regions, one of strong convection with low level convergence and upper level divergence, and a second region with opposite circulation, downward movement and dry conditions \citep{WH04,kiladis2014,vitart2017}. During its progression, MJO influences other phenomena like the Indian, Australian and South American monsoon systems and interacts with other phenomena such as ENSO \citep{kiladis2005, kim2009, fernandes2023}. Moreover, its influence is not limited to the tropics, it also impacts the extratropics through atmospheric teleconnection patterns.  During austral summer the MJO influences the South American rainfall Dipole (SAD), a rainfall pattern characterized by opposite rainfall anomalies in the South Atlantic Convergence Zone (SACZ) and southeastern South America (SESA). For example, during phases 8 and 1 (3 and 4) of MJO the SACZ (SESA) is in a wet phase and conversely for SESA (SACZ) (e.g. \citealt{alvarez2016, barreiro2019, grimm2019}). Both tropical-tropical and tropical-extratropical pathways are involved in this connection, which control the regional low level jet and thus the moisture transport, as well as the upper level circulation (e.g. \citealt{diaz2022}).

ENSO is the most important source of interannual climate variability, develops in the equatorial Pacific but has a strong influence around the globe \citep{garcia2006, shimizu2017, yeh2018, cai2020}.
In particular in South America during austral summer El Niño phase influences the rainy season in the coast of Peru, northeastern Brazil, southeastern Brazil, Uruguay, Argentina and Chile  (e.g. \citealt{ambrizzi2004, barreiro2010, cai2020, barreiro2024}). The relation between MJO and ENSO is a complex one  as they both influence the atmospheric circulation in the tropics. For example, MJO generates westerly wind bursts that last several weeks and that contribute to the development of El Niño \citep{fedorov2002, liang2021}. At the same time, some studies have shown that ENSO influences MJO through the modification of the center of enhanced and suppressed convection modifying its characteristics including the eastward propagation along the equator \citep{feng2015, fernandes2023, diaz2023}. \cite{chen2015} reported that the intensity of MJO is reduced during an El Niño event.

In a recent work \cite{corral2023} studied the statistics of the size and duration of MJO events and found that there is an increase probability of extinction after about 27 days, roughly half the average time the MJO needs to complete a full cycle. Moreover, this analysis allowed to distinguish between weak MJO and extreme MJO events, which would be expected to have different impacts worldwide and particularly over South America, but has not yet been fully addressed.

Hence in this study we focus on the characterization of extreme MJO events, their evolution and influence in tropical and extratropical South America depending on the initiation phase. Following \cite{corral2023}, MJO events are considered extreme when they exceed a certain threshold in duration or size. Finding this threshold is not an easy task, as there exist different methods to do so. In this study we tackle this problem by fitting a power-law to the tail of the distribution of MJO event duration and sizes \citep{corral2023}. Moreover, we further analyze the influence of the different ENSO phases on the MJO propagation and impacts.

In section \ref{sec2} we present the dataset and methodology employed. Section \ref{sec3} deals with the statistical characterization of extreme MJO events based on duration and size. Section \ref{sec4} shows the climate anomalies associated with extreme events depending on the phase of initiation and explores how the ENSO phase modifies the evolution of the MJO and impacts over South America during austral summer. Summary and conclusions are presented in section \ref{sec5}.

\section{Datasets and Methods}\label{sec2}

The evolution of MJO is analyzed by means of the \cite{WH04}, Real\textendash time Multivariate (RMM) index taken from the Australian Bureau of Meteorology during the period of 1979-2021 (43 years). This index is constructed using the first and second empirical orthogonal functions (EOF) considering  latitudinal averages (15°S\textendash 15°N) of outgoing longwave radiation (OLR) and zonal winds at 200 hPa and 850 hPa. The index provides the amplitude, or intensity of the MJO given by $ A = \sqrt{RMM1^2+RMM2^2}$ and its phase given by $\phi = \arctan(RMM2/RMM1)$, where RMM1 and RMM2 are the time series associated with the leading EOFs. The index comes with daily resolution.

\begin{figure}[ht!]
 \centering
 \captionsetup{margin=2.0cm}
 \includegraphics[width=0.3\textwidth]{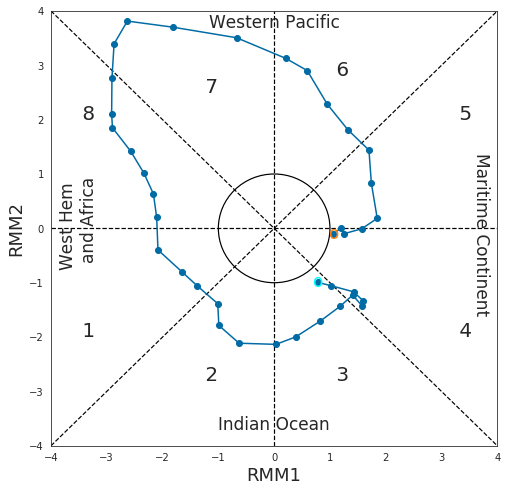}
 \caption{\footnotesize{Phase diagram, MJO progression from 03-03-2015 to 09-04-2015. The orange circle represents the first day of the event and the cyan circle the last day. Each blue point represents a day.}}\label{phdiag}
 \end{figure}

The index can be visualized in a phase diagram with RMM1 in the x-axis and RMM2 in the y-axis. The phase is discretized in eight values and the phase diagram is divided in terms of the phase in 8 geographical regions which are associated to the locations where the convection center is located. The western Hemisphere and Africa correspond to phases 8 and 1, the Indian Ocean to phases 2 and 3, the Maritime Continent (MC) to phases 4 and 5 and the western Pacific to phases 6 and 7, see Fig \ref{phdiag}. The eastward progression of MJO is represented in the phase diagram by a counterclockwise movement.

In our framework the MJO is active when the index amplitude is above or equal to 1 (outside the black circumference in Fig. \ref{phdiag}). Therefore, the contiguous time periods with active MJO define an event. Given this, we define the duration of an event in days as the period in which the amplitude is larger than 1 and the size of the event (its energy, roughly speaking) as the sum of the amplitudes along the duration, i.e.  $\sum_{t=t_i}^{t_f} A(t) $ as detailed in \cite{corral2023}.

For analyzing MJO events, in this study we use the outgoing longwave radiation (OLR; \cite{liebmann1996}) interpolated dataset from the National Oceanic and Atmospheric Administration (NOAA) which consist of daily data with a resolution of 2.5\grad x 2.5\grad\, for the period from 1979 to 2021. The winds dataset from the ERA5 reanalysis at 200 hPa is available at hourly resolution with a 0.25\grad x 0.25\grad \,spatial resolution, and has been preprocessed to daily and 2.5\grad x 2.5\grad \,to make it compatible with the OLR. The anomalies are obtained by subtracting the daily climatology.

The winds were used to calculate the streamfunction ($\psi $) and velocity potential ($\chi $) fields to describe the rotational and divergent components of the circulation, respectively. For the computation of these fields the windsparhm python library \citep{dawson2016} was used. 
The streamfunction allows a clear characterization of the tropical-extratropical teleconnections of the MJO mediated by Rossby waves \citep{straub2003, baxter2017, diaz2022}, and the upper level atmospheric response to enhanced convection which consists in a pair of anticyclonic circulation anomalies straddling the equator to the west of the heat source, and a pair of cyclonic anomalies to the west of the center of decreased rainfall. To better analyze these features we compute the eddy-streamfunction which is obtained by subtracting the daily zonal mean to the daily streamfunction.

The velocity potential is a scalar field that describes the divergent irrotational part of the velocity field. While the OLR describes the regional characteristics of the divergence, the velocity potential at 200 hPa describes the upper-level planetary-scale characteristics \citep{ventrice2013}. Moreover, the divergent wind is calculated as the gradient of the velocity potential $\chi$ given by $ \Bar{V} = \nabla \chi $.

Thus, negative velocity potential at 200 hPa means convergence in low levels with rising motion and divergent winds at 200 hPa. Positive velocity potential means convergence at 200 hPa and subsidence.

\subsection{Defining extreme events}

To study the progression of extreme MJO events we first need to estimate the optimal threshold above which an event is considered an extreme. As mentioned before we do this by fitting a power-law to the tail of the distribution of the duration and size of MJO events.

The probability density of a power-law distribution is:

\begin{equation}
f (x) =  \frac{\alpha}{a}\, \left( \frac{a}{x}\right)^{\alpha+1}\,\,\,   \textrm{for} \,\, x \geq a \label{pl} 
\end{equation}

and 0 otherwise, with $\alpha > 0$ and $a > 0$, where $\alpha + 1$ is the exponent of the density and $\alpha$ the exponent of the complementary cumulative distribution function and $a$ the lower cut-off. The variable $x$ may represent the duration or the size of MJO events.

We apply the Clauset et al's method and the Deluca’s method \citep{Clauset2009, Deluca2013} to the event duration and size distributions.
Both methods fit the tail of the distribution to a power-law by means of maximum likelihood and the  Kolmogorov-Smirnov goodness-of-fit test, using Montecarlo simulations \citep{Clauset2009, Deluca2013, Corral2019}. These methods allow one to find a lower cut-off or threshold for extremes.

\begin{figure}[ht!]
    \centering
      \includegraphics[width=0.85\textwidth]{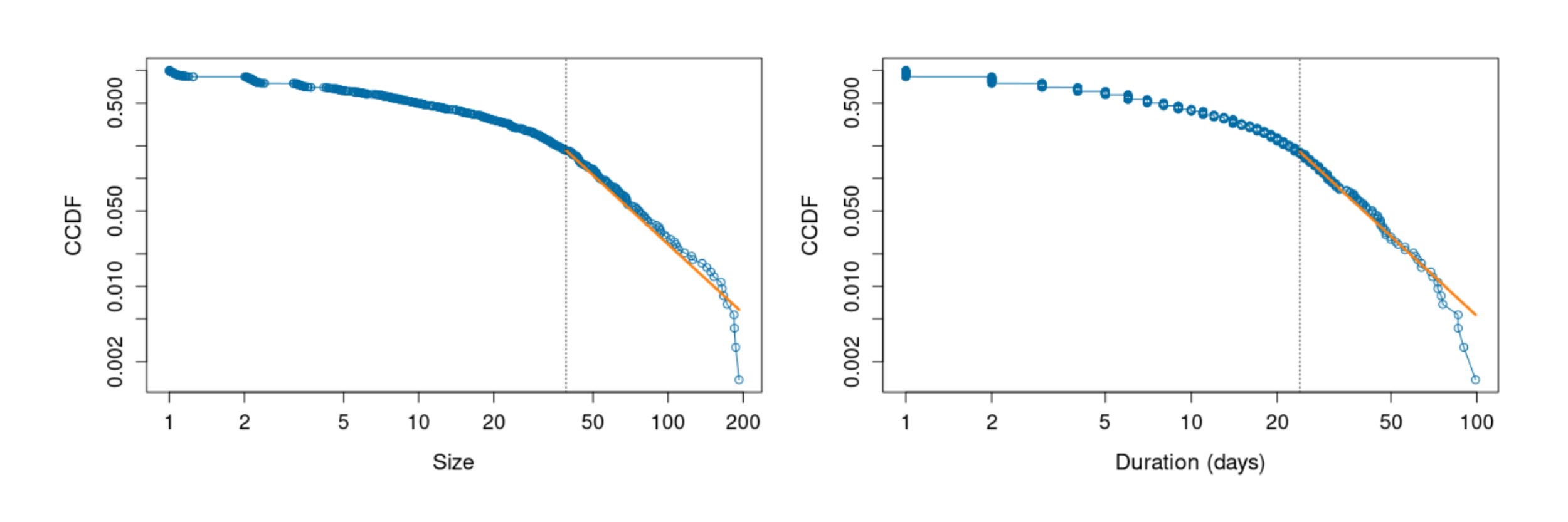}
        \caption{\footnotesize{Complementary cumulative distribution function (CCDF) of MJO events' size and duration. The orange line represents the power-law fit and the vertical dashed line marks the threshold, 39 for size (no units) (left) and 24 for duration (days) (right). }}
        \label{sdist}
\end{figure}

Considering events during all year round we find for the duration that extreme events are the ones that exceed or equals 24 days and for the size extremes are the ones that exceed or equals a value of 39 (Fig. \ref{sdist}). This is obtained by applying Clauset et al's method which yields results similar to the ones obtained with the Deluca's method. In this study we use the threshold based on size to define extreme events. Thus, according to this definition we call weak events the ones that have size smaller than 39 and further define non-events as the episodes consisting in all contiguous days with amplitude smaller than 1 and thus, MJO is inactive.

\subsection{Composite analysis}

The MJO evolution and its impacts can be identified using composite analysis of several events. In this study we aim at determining how the initial phase of extreme MJO events leads to different impacts on South America. The initiation of the MJO most commonly occurs over the Indian Ocean but the initiation is not limited to this area. For the composites we consider latitudes from 90\grad S to 20\grad N. 

We divide the initiation of events by pair of phases, that is, events that start in phases 8-1, 2-3, 4-5 and 6-7, and study the MJO impacts as it progresses through the following pair of phases until a cycle is completed or the MJO dissipates. We define a cycle when MJO passes through all of the phases. If MJO returns to the initial phase and continues its trajectory then we consider it a second cycle (or third). Each event is divided into different cycles, in order to avoid mixing days in the composites of the initial phases. We analyze only the progression over the first cycle as only few events continue after completing one.

When the amplitude is smaller than 1 (non events), the MJO is not active  and thus phases are not defined. In order to better quantify the impact of extreme events, the mean of all non events is subtracted from the composite of extremes for each pair of phases. The statistical significance of the composite of extreme events is then calculated based on these two populations using a two sample difference t-test at 10\% level. An analogous procedure is done for the composites of weak events.

Lastly, we construct composites of extreme MJO events based on the ENSO phases (El Niño, La Niña  and Neutral years) in order to determine the ENSO modulation of MJO’s impacts. ENSO phases are considered according to the ONI index taken from the Physical Sciences Laboratory of NOAA.

\vspace{0.5cm}

\section{Statistical characterization of extreme events}\label{sec3}

After estimating the optimal threshold, we compute the number of extreme events (size higher than or equal to 39), weak events (size smaller than 39) and non-events (days in which the index amplitude is smaller than 1), we summarize this information in Table \ref{evs1}. Regarding extreme MJO events we find 132 events from which 61 events start in extended austral summer (NDJFM) and 49 in extended austral winter (MJJAS). As expected weak events are more common with a similar occurrence in both seasons. Proportionally, during summer extreme events are more frequent than weak events.

\begin{table}[ht!]
\centering
\captionsetup{margin=2.5cm}
\caption{Number of extreme, weak and non events, for the whole year, extended summer and extended winter.}
\label{evs1}
\begin{tabular}{@{}lccc@{}}

\toprule
  & Total events & summer events  & winter events  \\ \midrule
Extreme events  & 132 & 61  & 49   \\
Weak events & 602  & 239  & 258   \\
Non events & 735 & 294  & 310   \\\bottomrule
\end{tabular}
\end{table}

\newpage
Looking more closely at extreme events, for the whole year we find significant more events starting in phases 2-3, while MJO initiation in phases 8-1, 6-7, and 4-5 only differs by 10\% (see table \ref{evs2}). During winter extreme MJO events also initiate mainly in phases 2-3, but during summer initiation occurs similarly starting in phases 2-3 and 8-1. On the other hand for weak events during the whole year we find considerably more events starting in phases 4-5, then in 2-3 and similarly in phases 8-1 and 6-7. During summer there are also more events starting in phases 4-5, and during winter initiation occurs similarly in phases 8-1 and 4-5 (see table \ref{evs3}).

\begin{table}[ht!]
\centering
\captionsetup{margin=2.5cm}
\caption{Number of extreme events starting in each pair of phases for all year round, extended summer and extended winter.}
\label{evs2}
\begin{tabular}{@{}lcccc@{}}

\toprule
Season  & Phases $8-1$ & Phases $2-3$ & Phases $4-5$ & Phases $6-7$ \\ \midrule
Whole year  & 32 & 42  & 27 &  31  \\
Summer & 18 & 19  & 11 &  13   \\
Winter & 8  & 18  & 11 &  12 \\\bottomrule
\end{tabular}
\end{table}

\begin{table}[ht!]
\centering
\captionsetup{margin=2.5cm}
\caption{Same as in table \ref{evs2} but for weak events.}
\label{evs3}
\begin{tabular}{@{}lcccc@{}}

\toprule
Season  & Phases $8-1$ & Phases $2-3$ & Phases $4-5$ & Phases $6-7$ \\ \midrule
Whole year  & 135 & 159  & 175 & 133   \\
Summer & 41 & 65  & 78 &  60   \\
Winter & 72 & 66  & 71 &  49 \\\bottomrule
\end{tabular}
\end{table}

Although the statistical characterization of duration and size of extreme MJO events was performed for austral winter and summer, the rest of the study is focused on the evolution of these events during austral summer when the MJO has the largest influence on South American rainfall (e.g. Alvarez et al. 2016).

\begin{figure}[ht!]
        \centering
        \sbox0{\includegraphics[width=0.8\textwidth]{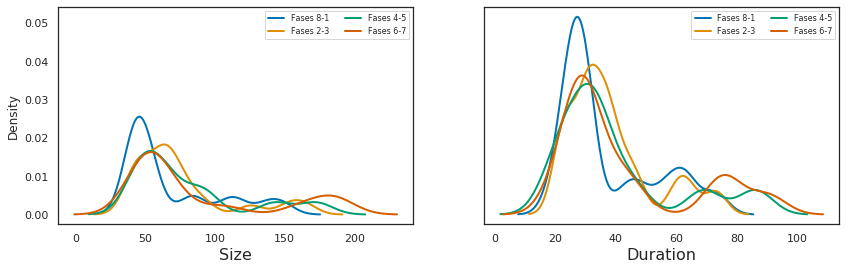}}
        \begin{minipage}{\wd0} 
        \usebox0 
        \caption{\footnotesize{Distributions of extreme MJO events duration and size for each initial pair of phases during austral summer. The area  under the curve is 1, a Gaussian function was used as a kernel function to estimate the density. The densities should be zero at 39 for the size and 24 for the duration (threshold for each variable) but the smoothing makes the density $\neq 0$ in this points.}}\label{f1}
        \end{minipage}
\end{figure}

In figure \ref{f1} we plot the distributions of the duration and size of the extreme events that start in each pair of phases in austral summer. We can observe that MJO events starting in phases 8-1 have shorter durations and size than the rest of the phases. On the contrary, MJO events starting in phases 2-3 tend to last longer and have larger sizes.  
Both sets of distributions present a long tail suggesting that there is a lot of variability among extreme events in terms of duration and size.

\begin{figure}[ht!]
        \centering
        \captionsetup{margin=2.0cm}
        \includegraphics[width=0.8\textwidth]{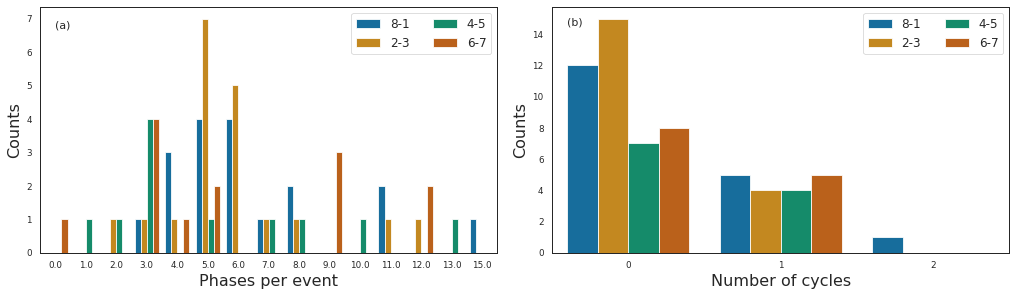}
        \caption{\footnotesize{(a) Phase trajectory distribution showing the number of phases traversed by each extreme event during austral summer, colored by initial pair of phases and (b) count distribution of MJO cycle completions during austral summer colored by initial pair of phases.}}
        \label{f2}
\end{figure}

In addition to the size and duration of an MJO event we looked at the advance in phases for extreme MJO events starting in each phase (Fig. \ref{f2}a). Zero advance means that an event starts and finishes in the same phase. Events that propagate along 8 phases or more are events that continue after completing a cycle.  Clearly, most extreme MJO events do not continue past a complete cycle, but several events do, and there is a dependence on the initial phase and season of the year. 
Extreme MJO events starting in phases 2-3 have a clear maximum in propagation of 5 phases, while those starting in phases 8-1 tend to propagate for 5 or 6 phases. On the other hand, extreme MJO events that start in phases 4-5 tend to progress over 3 phases and those starting in phases 6-7 do not have a clear maximum in propagation phases. According to Figures \ref{f1} and \ref{f2} events that start in phases 8-1 have smaller size and duration but they propagate faster compared to events starting in phases 2-3.

When we look closer to the events that have a progression of zero, one and two phases we see that the event with zero advance has a duration of 32 days, the event with one phase of advance has a duration of 23 days and the events that progress over two phases have a duration of 23 and 32 days. This is not rare as we are looking at extreme events and the shorter event has a duration of 19 days. 
Lastly figure \ref{f2}b shows the number of cycles completed by extreme MJO events. We can observe that few events continue after the first cycle is completed and only one event finishes a second cycle.

\section{Composite analysis}\label{sec4}

\subsection{Impacts of extreme and weak events}

We apply a composite analysis to look at the evolution of extreme MJO events during summer depending on the phase of initiation and how they influence remote regions, particularly South America and the SAD. Although we analyzed the evolution of events starting in every phase, we focused mainly on events starting in phases 2-3 and 6-7 since they show the strongest anomalies over South America according to our results.

\begin{figure}[ht!]
    \centering
    \captionsetup{margin=2.0cm}
        \includegraphics[width=1.0\textwidth]{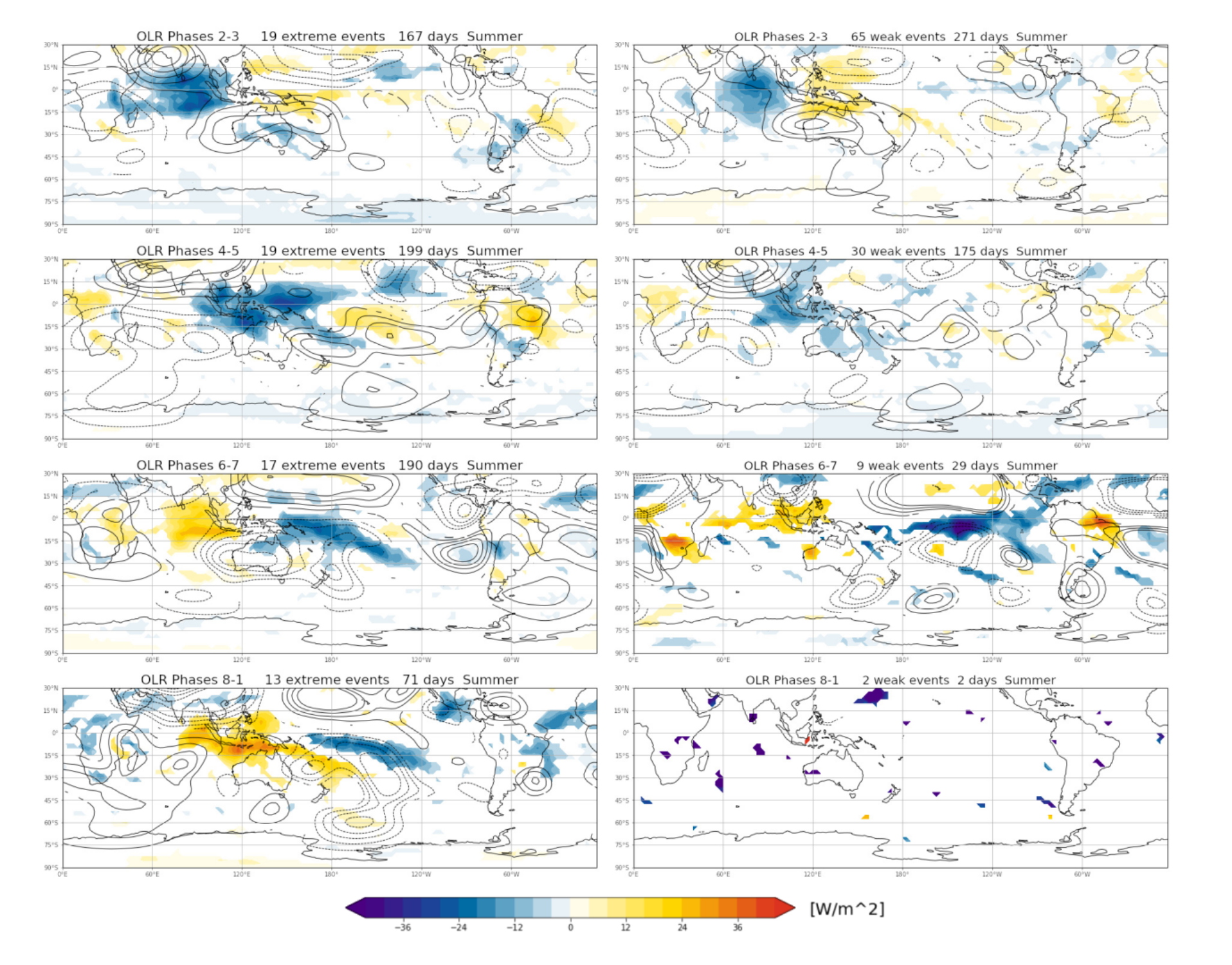}
      \caption{\footnotesize{Composite maps of OLR ($W m^{-2}$) anomalies of (left column) extreme events, (right column) weak events and 200 hPa eddy streamfunction contours with intervals of $2.0\textrm{x}10^{-6} m^2 s^{-1} $. Only 90\% significant values are shown. Positive values (straight lines) represent a cyclonic gyre and negative values (dashed lines) represent anticyclonic gyre in the southern hemisphere. Each column shows the evolution of events starting in phases 2-3.}}\label{f4} 
    
\end{figure}

\begin{figure}[ht!]
    \centering
    \captionsetup{margin=2.0cm}
        \includegraphics[width=1.0\textwidth]{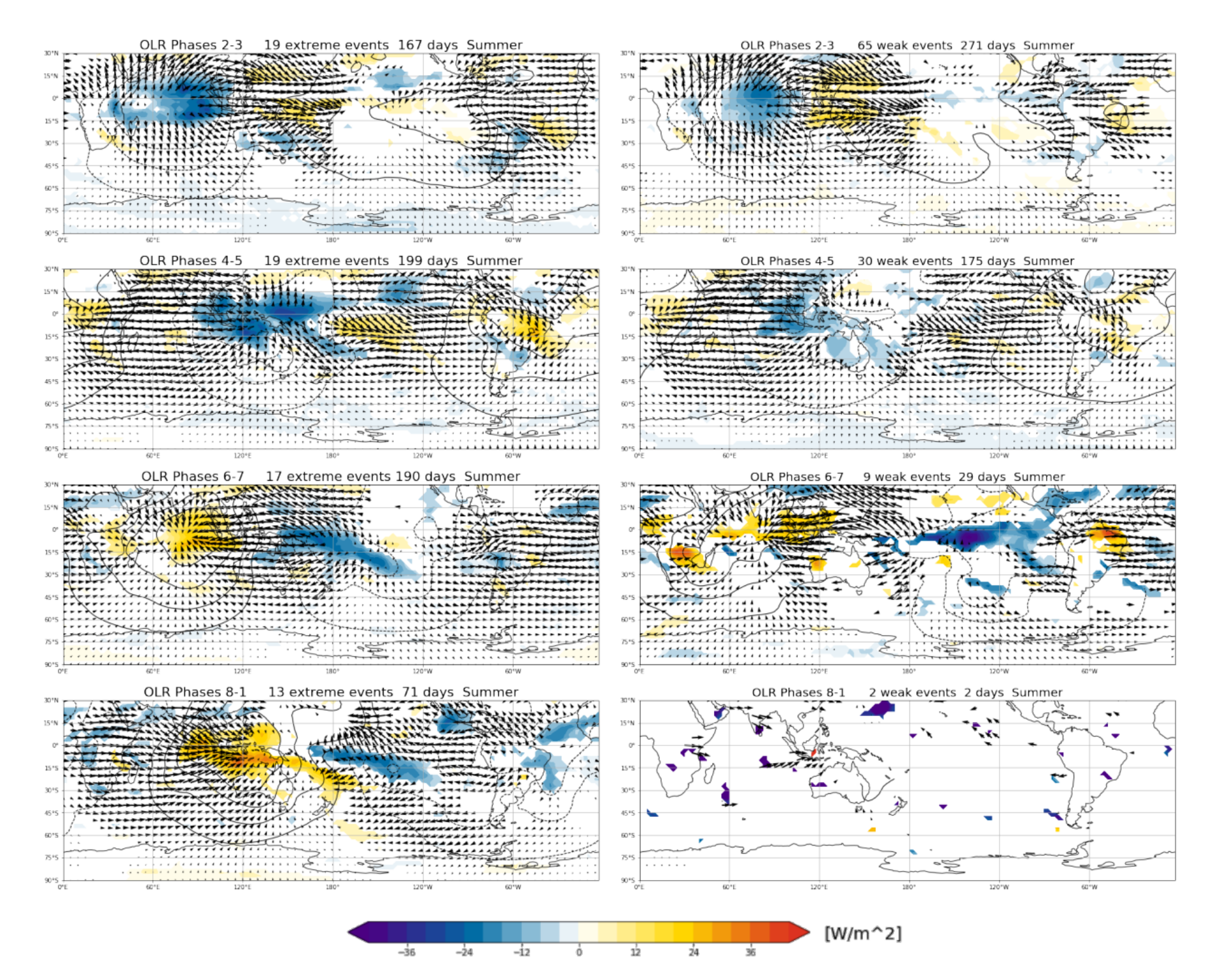}
      \caption{\footnotesize{Composite maps of OLR anomalies of (left column) extreme events, (right column) weak events and 200 hPa velocity potential contours with intervals of 1.5x$10^{-6} m^2 s^{-1} $. Positive values (straight lines) follow the center of suppressed convection and negative values (dashed lines) follow the center on enhanced convection. The arrows represent the divergent wind, only 90\% significant values are shown for all variables.  Each column shows the evolution of events starting in phases 2-3.}}\label{f5} 
    
\end{figure}

\begin{figure}[ht!]
    \centering
    \captionsetup{margin=2.0cm}
        \includegraphics[width=1.0\textwidth]{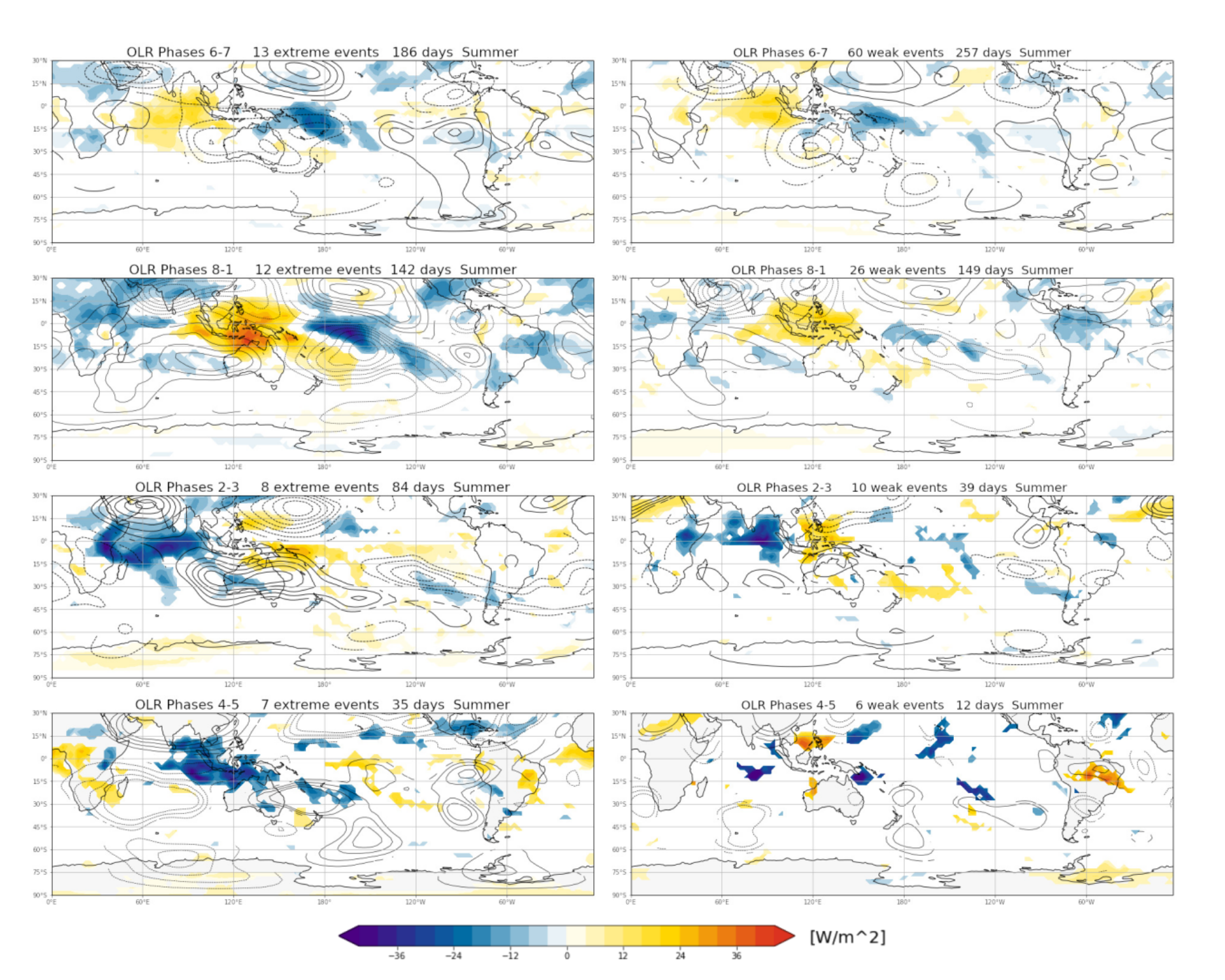}
       \caption{\footnotesize{Same as in fig \ref{f4} but for events starting in phases 6-7.}}\label{f6}
    \end{figure}

\begin{figure}[ht!]
    \centering
    \captionsetup{margin=2.0cm}
        \includegraphics[width=1.0\textwidth]{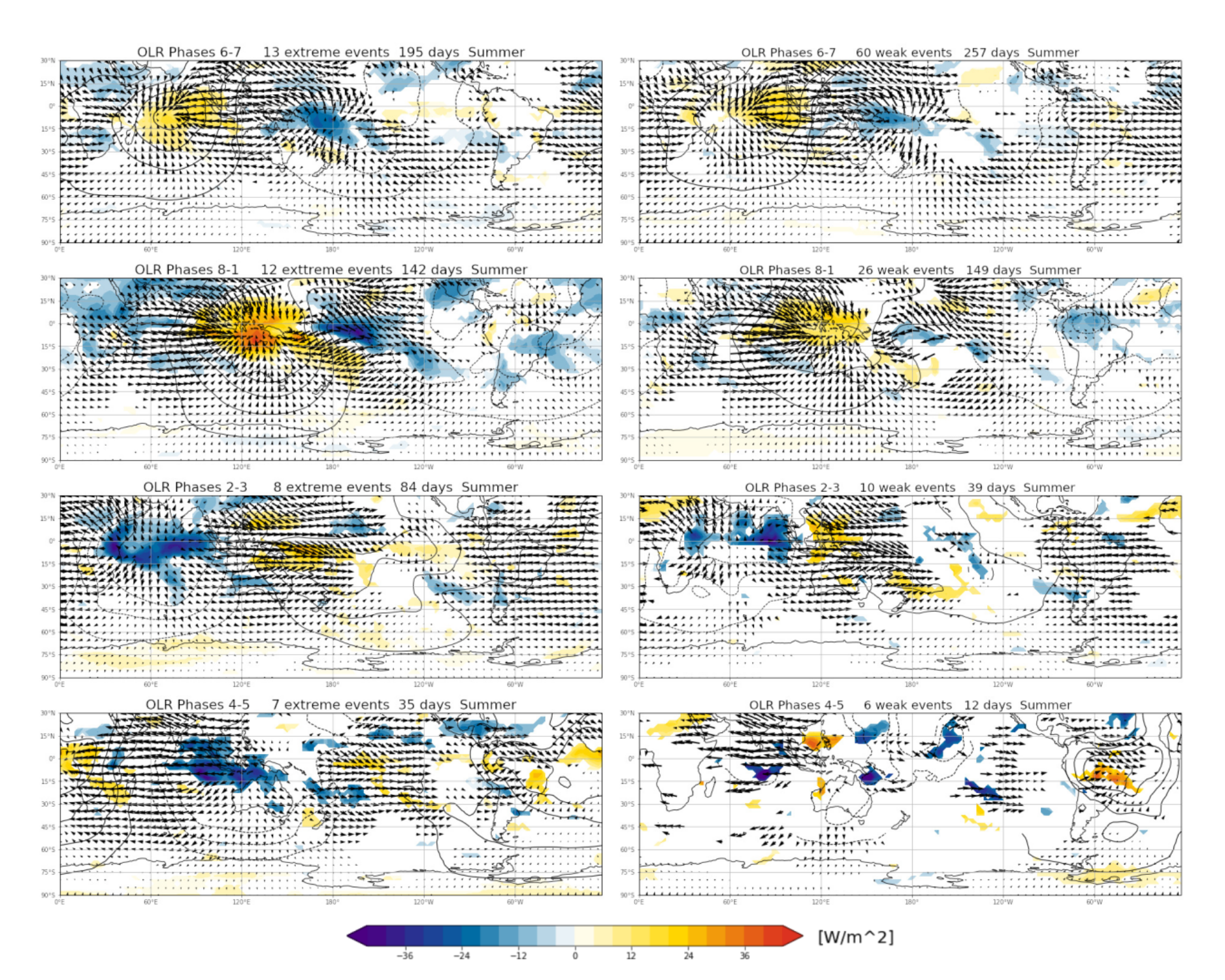}
       \caption{\footnotesize{Same as in fig \ref{f5} but for events starting in phases 6-7.}}\label{f7}
    
\end{figure}

Figures \ref{f4} to \ref{f7} illustrate the evolution of extreme events (left column) and weak events (right column), beginning from each initial pair of phases and tracking their progression through the subsequent phase pairs. For instance, in Figure \ref{f4}, the top panel displays the composite of all events that originated in phases 2-3 while the events stayed in phases 2-3. The second panel shows how these events evolved into phases 4-5, followed by their progression into phases 6-7 in the third panel, and finally into phases 8-1 in the bottom panel. Each panel indicates the number of events that advanced to that phase pair, as well as the total number of days the events remained within each phase pair. 

In Figure \ref{f4} each panel shows the anomalies of OLR and eddy-streamfunction. Clearly, OLR anomalies are stronger in extreme events compared to those in weak events. For weak events during the initial phases 2-3, negative OLR anomalies (increased precipitation) are seen in the north of South America and positive anomalies (dry conditions) in eastern Brazil. Both cases, extreme and weak events, show increased precipitation over southern Brazil, Uruguay and Argentina but it can be seen that during extreme events the anomalies are more intense and extend further south over Argentina and also to Chile. The positive OLR anomalies over eastern Brazil (in the SACZ) and the negative anomalies in southeastern South America conform the SAD. In the extreme events case, the circulation presents upper-level cyclonic-anticyclonic anomalies located over subtropical South America evidencing the existence of a trough, which favors dynamic lift over SESA thus promoting rainfall there and at the same time decreasing rainfall over the SACZ. Only the anticyclonic anomaly is seen in the weak events composite. 
Negative OLR anomalies can also be seen over Australia for extreme events which are not seen on weak events until the next pair of phases (4-5).

As the MJO moves to phases 4-5 the center of enhanced convection in weak events is located over Sumatra, while in extreme events there is an additional and stronger center located in the western Pacific. Anomalies in the weak composite only show a weak rainfall dipole over eastern South America with anomalies of similar magnitude as in phases 2-3. On the contrary, in the extreme events case the dipolar rainfall anomalies now cover a larger extent including tropical South America. The subtropical anticyclone has weakened but still plays a role in increasing rainfall over Uruguay. In addition, the divergent wind clearly shows strong convergence over equatorial South America as a consequence of the MJO displacement to the east (Figure \ref{f5}). This tropical-tropical teleconnection increases the subsidence over the Amazon region, thus decreasing rainfall. As a consequence the rainfall dipole increases in extent due to the tropical-tropical and tropical-extratropical teleconnections.

Some authors have pointed out the barrier effect of the MC over the eastward propagation of MJO as most of the MJO events do not propagate through the MC or become weakened \citep{wu2009, zhang2017, ling2019, barrett2021}. However, it can be observed in Figure \ref{f4} that this does not seem to be the case for extreme events starting in phases 2-3 as more than 68\% of the events that reach phases 4-5 propagate beyond the MC. The same happens for events starting in phases 8-1 and 4-5 (not shown). In contrast, only few weak events that reach phases 4-5 propagate beyond the MC (regardless of the phase of initiation), in fact, more than 50\% of the weak events that start in any phase are lost before reaching the next phase (see Figures \ref{f4} and \ref{f6}).

In the third row of Figure \ref{f4} (phases 6-7) it can be seen that only 9 weak events (with a total of 29 days) propagate beyond the MC, the composite is noisy and does not represent correctly the evolution of MJO. On the other hand almost all extreme MJO events reach phases 6-7. The SAD has almost disappeared and now the largest negative OLR anomalies are located in northern Chile, Bolivia and Peru.

When extreme MJO events reach phases 8-1 the resulting anomalies over South America are similar in structure and opposite in sign to those during phase 4-5: a cyclonic anomaly develops off Uruguay which favors rainfall in the SACZ. Velocity potential and divergent anomalies in phases 8-1 are also opposite to those in phases 4-5 (Figure \ref{f5}). It is worth mentioning that there is a drop in the number of extreme events and particularly in the number of days that make up the composite for phases 8-1, in agreement with Figure \ref{f2}a. Nevertheless, it is still much larger than the number of weak MJO events that make it to phases 8-1.

The results presented here for the extreme MJO events are in agreement to what \citealt{diaz2022} reported on the evolution of long and intense MJO events starting in phases 2-3 from day zero to day 20.

\newpage

We now study the evolution of MJO events when they start in phases 6-7. As mentioned in Table \ref{evs1}, extreme MJO events are fewer than those starting in phases 2-3, but the number of weak events does not change considerably. Convection and circulation anomalies associated with extreme MJO events are much more intense and located further east than those of weak MJO events Figure \ref{f6}. Over South America, for extreme MJO cases, there are negative OLR anomalies in Chile and Argentina, located south of the anomalies seen in phases 6-7 when the MJO started in phases 2-3 (compare Figure \ref{f4} and \ref{f6}).

These anomalies become stronger and larger in extent as the MJO progresses to phases 8-1. In addition, large negative OLR anomalies develop over the SACZ, as also seen in Figure \ref{f4}. On the other hand, in weak events large negative OLR anomalies develop in northern South America in phases 8-1. This difference in rainfall response can be traced back to the development of a strong convective center in central Pacific in the extreme MJO case, which induces upper level convergence over northern South America. In the case of weak MJO events, convection in central Pacific is weak and the velocity potential shows a smoother wavenumber 1 structure with centers in northern South America and the MC in Figure \ref{f7}.

For extreme MJO cases, when the convection center reaches phases 2-3 the negative OLR anomalies situated in Chile and Argentina have diminished and there is no sign of the SAD, an important difference with those extremes that initiate in phases 2-3. As the MJO moves to the last pair of phases (4-5), there are only few days left to represent the effects of MJO in the continent but for both weak and extreme cases there is a tendency to develop the SAD mainly through a decrease in rainfall over the SACZ. Positive anomalies of the velocity potential over that region suggest a tropical-tropical teleconnection (Figure \ref{f7}).

Thus, comparison of the evolution of extreme MJO events starting in phases 2-3 and 6-7 shows that OLR anomalies develop over some regions of South America irrespective of the initiation phase, while in other regions this is not the case. This suggests larger subseasonal predictability in the former regions.

\newpage

\subsection{Influence of ENSO on MJO}

\begin{figure}[ht!]
        \centering
        \captionsetup{margin=2.0cm}
        \includegraphics[width=0.80\textwidth]{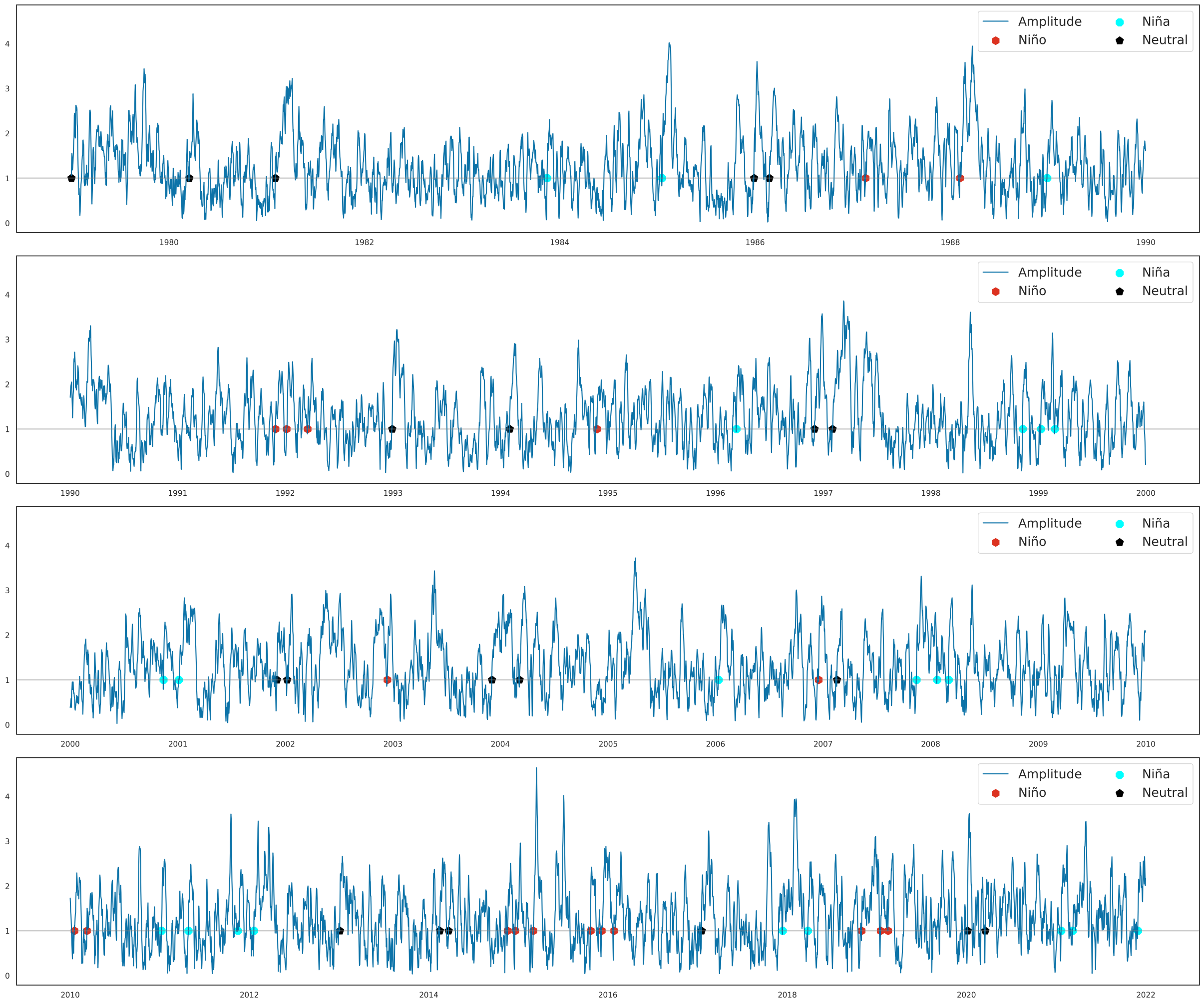}
        \caption{\footnotesize{Temporal evolution of the MJO index amplitude (blue) from 1979 to 2022, overlaid with coincidences of ENSO phases. Markers indicate coincidences of austral summer MJO extreme events with El Niño (orange), La Niña (cyan), and Neutral (black) phases.}}\label{f9}
\end{figure}

To analyze whether ENSO influences the impact of extreme MJO events over South America we first categorized extreme MJO events based on the coincidence of their initiation time with El Niño (EN), La Niña (LN), or Neutral (NT) phases. Figure \ref{f9} shows the MJO amplitude (blue), with markers indicating occurrences where extreme MJO events coincide with ENSO phases: EN (orange), LN (cyan), and NT (black).

\newpage

\begin{figure}[ht!]
 \centering
 \captionsetup{margin=1.5cm}
 \includegraphics[width=0.6\textwidth]{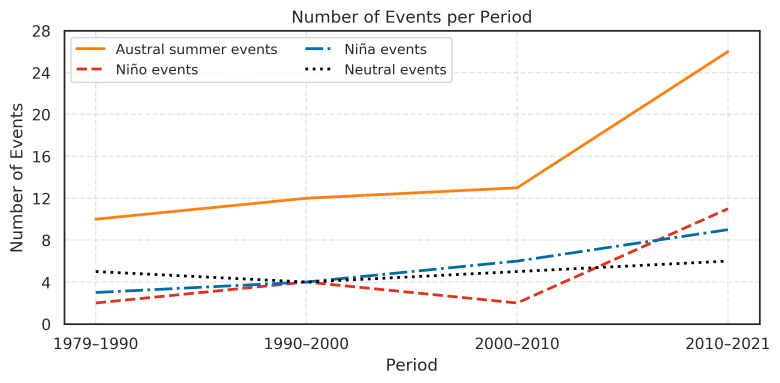}
 \caption{\footnotesize{Number of extreme MJO events (orange), along with those that occurred during El Niño (red), La Niña (blue) and Neutral (black) in four different periods during austral summer.}}\label{fig:extenso}
\end{figure}

Out of the 61 extreme MJO events that take place during summer, 19 coincide with EN, 22 with LN and 20 with the NT Phase. Interestingly, more than half of the extreme events during EN years occurred from 2010 to 2021. During LN years extreme events also increased but not as much as during EN  and during NT extreme events do not show a change in the different decades, see Figure \ref{fig:extenso}. As a consequence  during the last period (from 2010-2021) the number of extreme MJO events has increased considerably compared to the previous decades. Compared to the periods 1990–2000 and 2000–2010, the number of events during 2010-2021 approximately doubled, and it nearly tripled relative to the earlier period of 1979–1990. This increase of extreme MJO events is seen during austral summer, when MJO is strongest \citep{wang2020}, but not in austral winter (not shown).

\begin{figure}[ht!]
        \centering
        \sbox0{\includegraphics[width=0.63\textwidth]{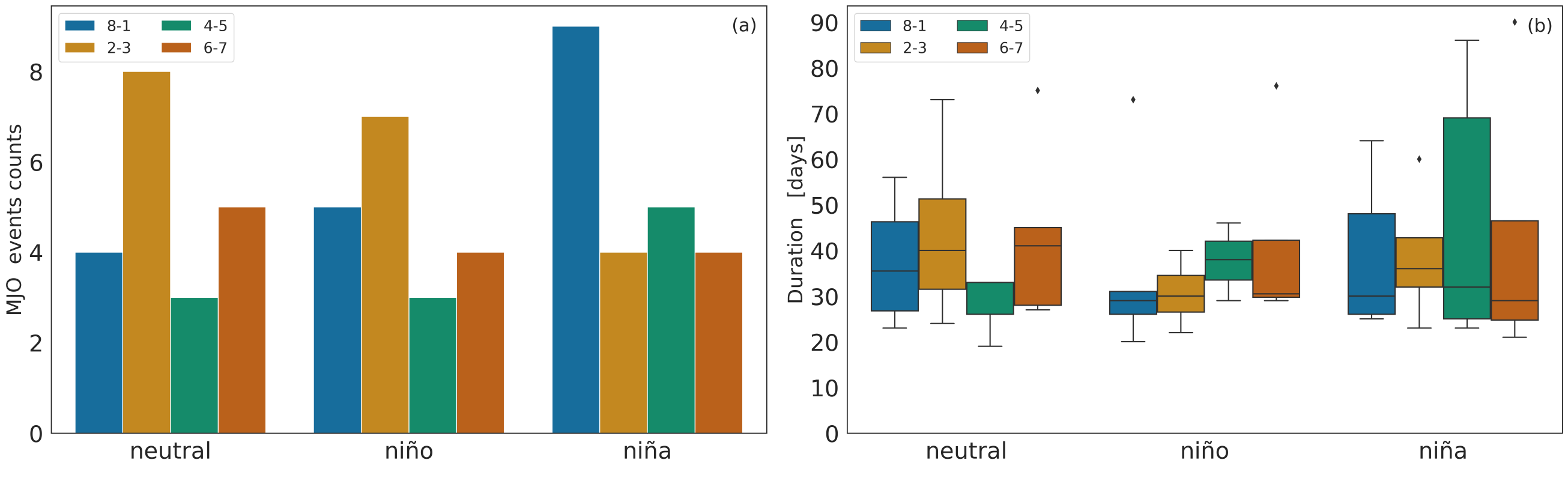}}
        \begin{minipage}{\wd0} 
        \usebox0 
        \caption{\footnotesize{a) Distribution of extreme MJO event counts associated with each ENSO state, categorized by phase of initiation. b) Distribution of event durations associated with each ENSO state, categorized by phase of initiation.}}\label{box1}
        \end{minipage}
\end{figure}

\begin{figure}[ht!]
        \centering
        \sbox0{\includegraphics[width=0.63\textwidth]{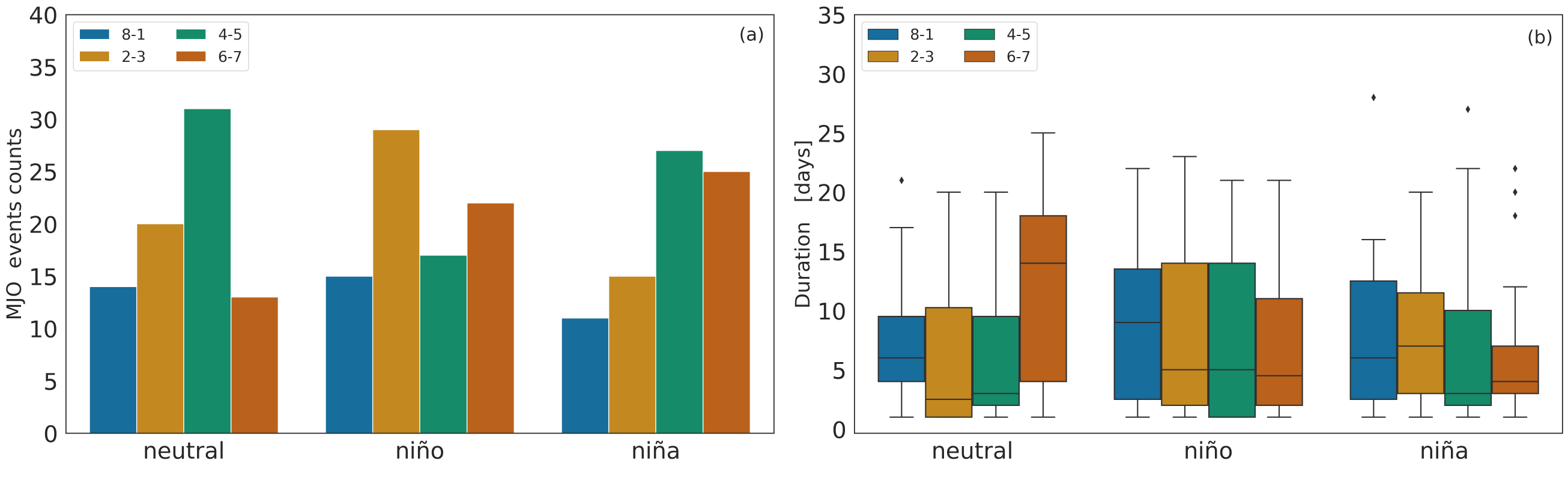}}
        \begin{minipage}{\wd0} 
        \usebox0 
        \caption{\footnotesize{Same as in Figure \ref{box1} but for weak events.}}\label{box2}
        \end{minipage}
\end{figure}

With respect to MJO initiation, we find that more extreme MJO events start in phases
8-1 during LN (Fig. \ref{box1}a), consistent with the findings of \cite{wang2022}. They reported that LN-like background patterns favors the initiation of strong MJO events in Atlantic and Africa (phases 8-1). On the contrary, we find that during EN extreme MJO events tend to initiate in phases 2-3, while they found that more events initiate in phases 4-5 and 6-7. For NT ENSO years there are more MJO events starting in phases 2-3, then in phases 6-7, 8-1 and last in phases 4-5. Figure \ref{box2}a reports the same statistics for weak events showing that they behave similarly during EN years but not during LN or NT years. For example, during LN years weak events tend to initiate more frequently in phases 4-5 and 6-7.

Regarding the lifetime of extreme MJO events we find that the durations of events
during NT and LN are longer than during EN, (with means $40\pm 15$, $40\pm 20$ and $35\pm 14$ days respectively) while the mean duration of weak events is similar in all ENSO phases ($8\pm 6$ for EN, $7\pm 6$ for LN and $7\pm 6$ for NT). These results are opposite to what \cite{shimizu2017} reported for all active MJO during the period of 1979-2010. Figures \ref{box1}b and \ref{box2}b show the mean duration of extreme and weak events as a function of the initiation phase.

\newpage

\begin{figure}[ht!]
    \centering
     \captionsetup{margin=2.0cm} 
        \includegraphics[width=0.50\textwidth]{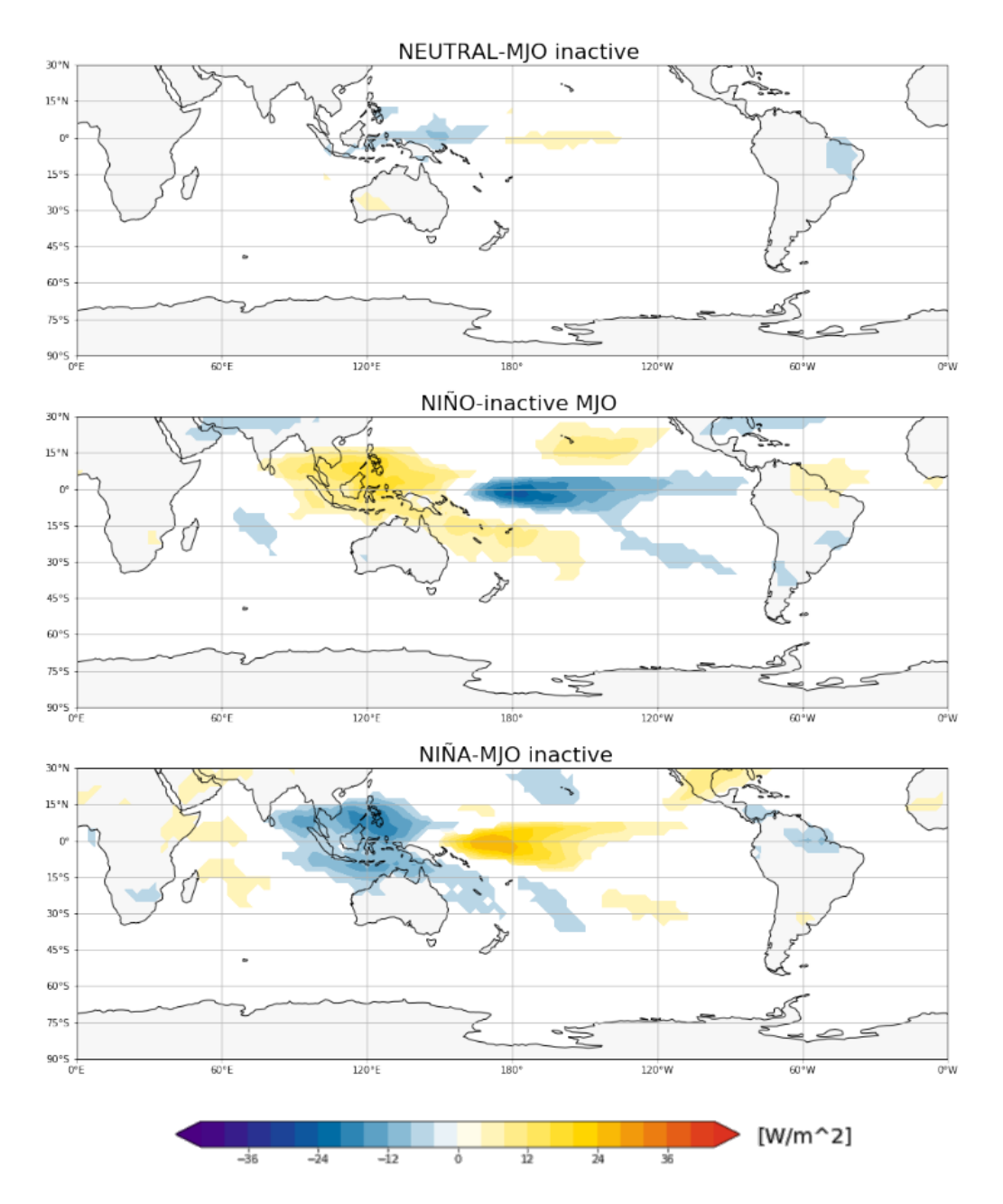}
     \caption{\footnotesize{Composites of OLR anomalies during NT, El Niño and La Niña conditions with inactive MJO (1979 - 2021). Only values significant at the 90\% confidence level are displayed.}}\label{onlyenso}
     \end{figure}

As a reference for the discussion below Figure \ref{onlyenso} shows the OLR anomalies during different ENSO states when the MJO is inactive. As expected, the OLR shows the typical shift in convective activity in the tropical Pacific with enhanced rainfall on the central Pacific and decreased rainfall over the MC during EN and vice versa for LN years. During NT years OLR anomalies are very weak.

 \newpage 
Next we separate the composites of extreme MJO events in events that coincide with each ENSO state, NT (left column), EN (middle column) and LN (right column). Given the location of OLR anomalies characteristic of EN, LN and MJO one would expect that Pacific OLR anomalies during MJO phases 4-5 tend to be in phase with LN and during phases 8-1 in phase with EN. Thus, climate anomalies over South America will be the result of the interference between these two phenomena that have different characteristic time scales.     

\begin{figure}[ht!]
    \centering
     \includegraphics[width=1.0\textwidth]{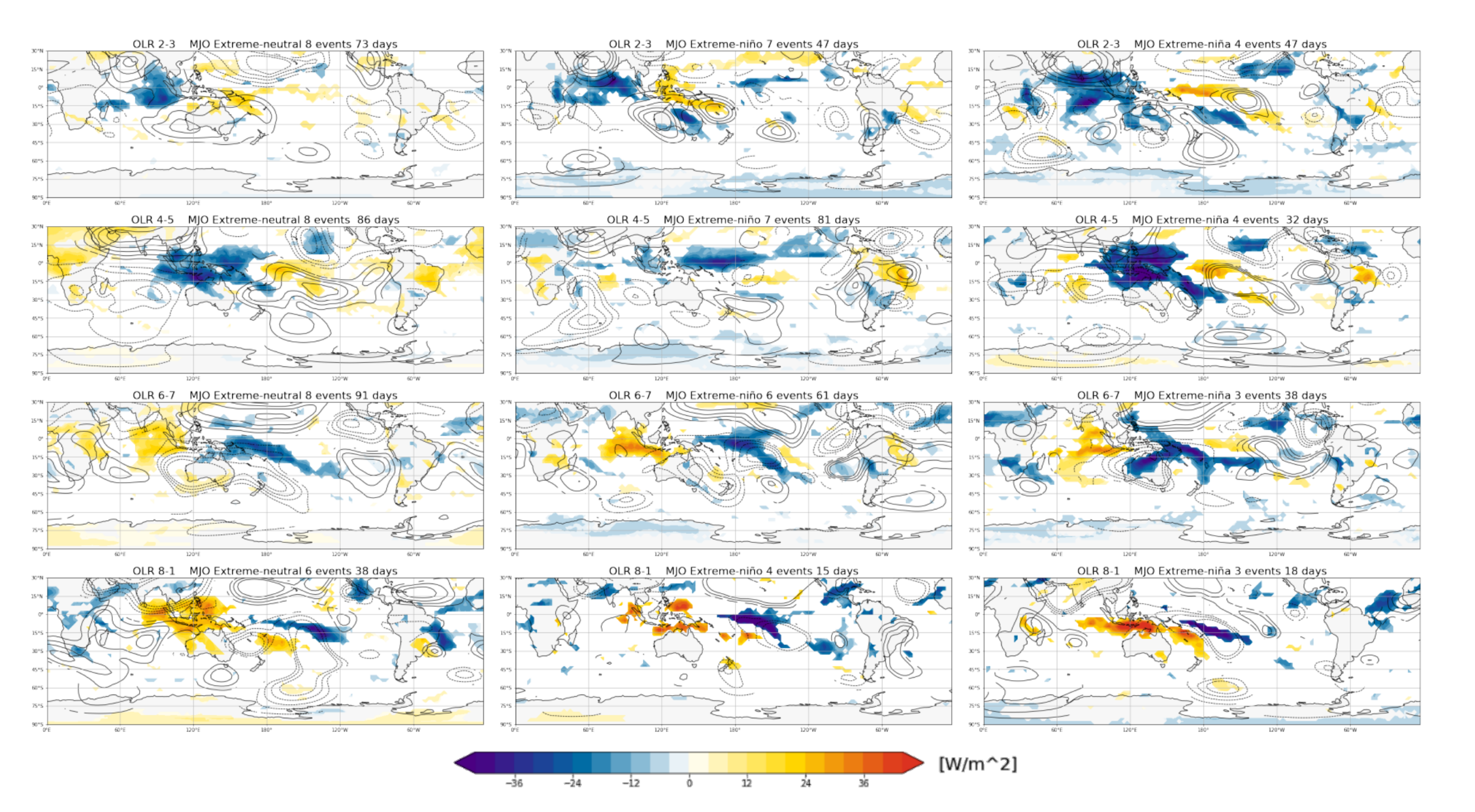}
     \caption{\footnotesize{Evolution of extreme MJO events that start in phases 2-3 and that coincide with (left column) NT ENSO phase, (middle column) el Niño phase and (right column) la Niña phase.Only significant values with confidence levels better than 90\% are shown.}}\label{enso23}
\end{figure}

For extreme MJO events starting in phases 2-3 there are more events during NT and EN phases compared to LN years but with more days for the NT state (Figure \ref{enso23}). NT events do not present significant precipitation anomalies over South America in phase 2-3, while during EN and LN it is clear the appearance of the SAD associated with cyclonic-anticyclonic circulation anomalies in upper levels, as shown in Figure \ref{enso23}, particularly during EN. As the MJO progresses, and reaches phases 4-5 the center of enhanced convection in the Indo-Pacific regions for NT and LN years is very similar to the MJO composite for all extreme cases. On the other hand, convective anomalies during EN have smaller extent and cover only the equatorial region as the OLR anomalies associated with the two phenomena interact destructively.

Nevertheless, in these phases (4-5) the SAD is clearly seen independently on the phase of ENSO and is strongest during EN due to the presence of larger circulation anomalies over South America. Moreover, although weakened, the rainfall dipole during EN persists up to phases 6-7, but does not continue in NT or LN. In particular, the dry center in eastern Brazil extends up to the equator in the EN case, probably because this phenomenon increases vertical stability over the Amazon region, while during LN rainfall increases there, see Figure \ref{onlyenso}. At the same time the negative OLR anomaly located in subtropical South America is larger during EN due to a stronger upper level anticyclonic circulation located off Uruguay, also characteristic of EN and not found in NT periods. 

Only a few extreme MJO events evolve up to phases 8-1 during LN and EN and thus the composites are noisy. However, during NT years there is still a large number of days and phases 8-1 show a clear SAD of opposite sign from that during phase 4-5, as also seen in Figure \ref{f4}.

\newpage

\begin{figure}[ht!]
    \centering
       \includegraphics[width=1.0\textwidth]{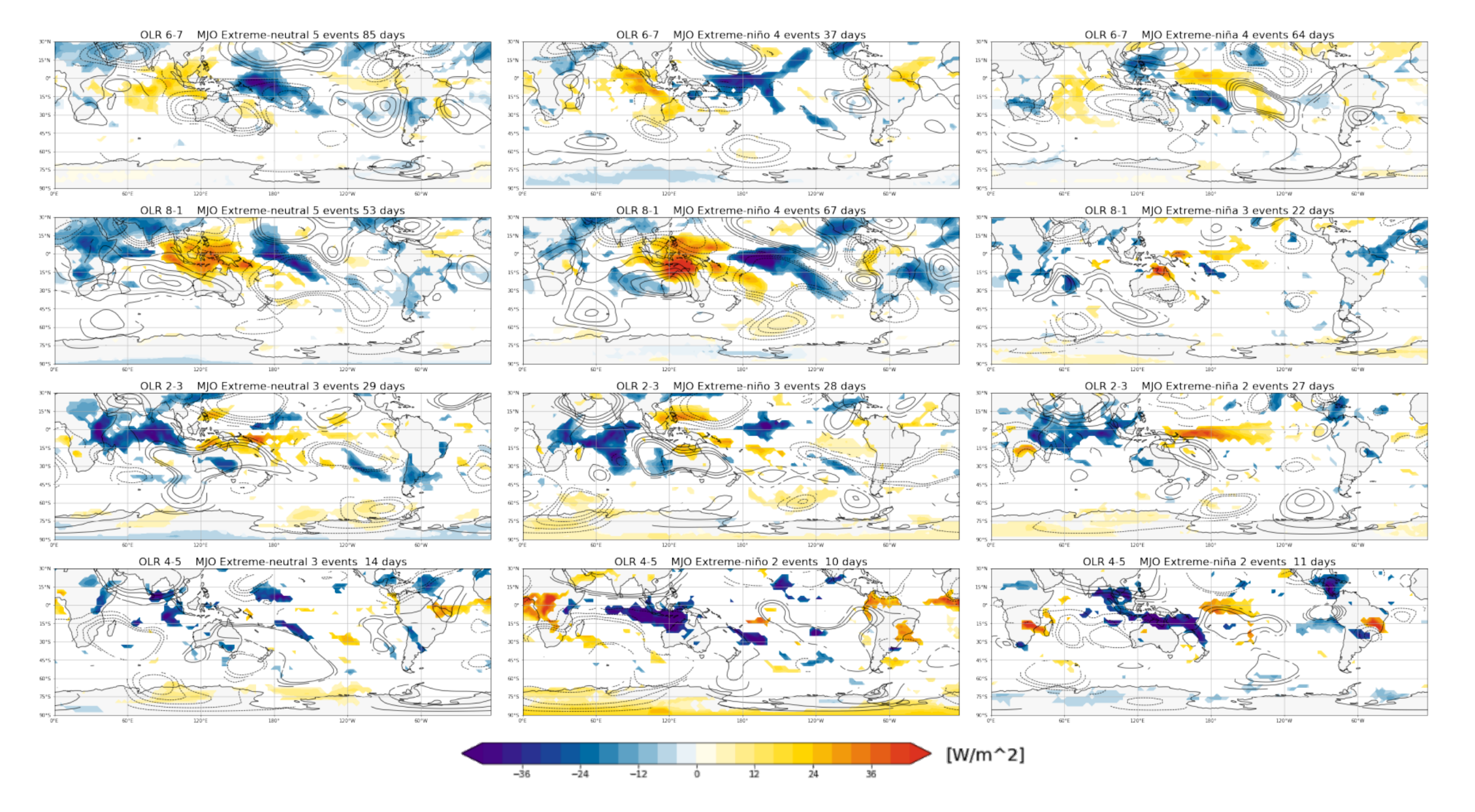}
     \caption{\footnotesize{Same as in \ref{enso23} but for events starting in phases 6-7. }}\label{enso67}
\end{figure}

Opposed to events starting in phases 2-3, when the enhanced convection starts over the east Pacific, phases 6-7, the composite of extreme MJO events during NT ENSO is the one that presents OLR anomalies that share largest similarities with the MJO composite of Figure \ref{f6}, although negative anomalies are further to the southeast (Figure \ref{enso67}). Only few MJO events evolve to phases 8-1 during LN years and the composites become noisy. During NT and EN years there are negative OLR anomalies over Argentina and Chile. However, only during EN years there is a clear enhancement of the SACZ possibly as consequence of constructive interference between the two phenomena. 

There is a strong decrease in the number of extreme MJO days that progress to phases 2-3 and composites become noisy for all ENSO phases. The most robust signal appears to be the negative OLR anomaly over Argentina and Chile during NT years, also present in Figure \ref{f6} albeit weaker.  For the last pair of phases (4-5) each composite is constructed with very few days and thus may not represent the behavior accurately.

\newpage

\section{Summary and Conclusions}\label{sec5}

This work studies the evolution of extreme Madden-Julian oscillation (MJO) events and its impact over subtropical and  extratropical South America during extended austral summer.  
We first find a threshold above which an event is considered extreme by fitting a power-law to the tail of the distribution of MJO events' size (given by the sum of the amplitudes along the duration of an event) \citep{corral2023}. 
According to this definition, extreme events are the ones with a size $\geq 39$ and weak events the ones with a size < 39. Extreme MJO events most frequently initiate in phases 2-3 throughout the year, with similar initiations across phases 8-1, 6-7, and 4-5. This distribution is also seen in  winter, while in summer, initiation is more balanced between phases 2-3 and 8-1. In contrast, weak events predominantly start in phases 4-5 year-round, followed by phases 2-3, with phases 8-1 and 6-7 occurring at similar frequencies. Seasonally, weak events continue to favor phases 4-5 in summer, while in winter, initiation is evenly distributed between phases 8-1 and 4-5.

Using composites of OLR, streamfunction and divergent wind anomalies we analyzed the progression of extreme MJO events considering the phase of initiation. We focus on climate anomalies over South America for each initiation pair of phases and particularly for events initiating on phases 2-3 and 6-7. 
We found that the centers of enhanced and suppressed convection during extreme MJO events are more intense and extend further east and south than those of weak MJO events.

Similarly, over South America extreme MJO events present more intense rainfall and circulation anomalies and are larger in spatial extent. During extreme events we can detect the tropical-extratropical teleconnections over Chile, Argentina, Uruguay and south Brazil while during weak events the tropical-tropical teleconnections predominate and the anomalies found over the subtropical South America are weaker and cover  smaller regions. 

Most extreme MJO events propagate through all of the phases completing a cycle and they are strong enough to propagate beyond the Maritime Continent (MC) regardless of the phase of initiation while only half of the weak events continue to the second pair of phases. This is consistent with the fact that weak events have short durations.

The South American rainfall Dipole (SAD), the main mode of variability of precipitation over South America, appears clearly in phases 2-3 for events starting in these phases when the center of enhanced convection is located over the Indian Ocean.  
During extreme events, the circulation presents upper-level cyclonic-anticyclonic dipolar anomalies located over subtropical South America leading to enhanced precipitation over SESA extending south over Argentina and decreasing rainfall over SACZ. As the center of enhanced convection moves to the MC (phases 4-5) the tropical-tropical teleconnection favors strong convection over equatorial South America thus increasing subsidence over the SACZ region hence increasing the rainfall dipole. The dipole almost disappears as MJO moves to the western Pacific (phases 6-7) and it finally changes in sign as the MJO reaches the western hemisphere (phases 8-1). 
During weak MJO events in phases 2-3 only the anticyclonic circulation anomaly develops off Brazil accompanied by weak positive OLR anomalies over eastern Brazil and negative anomalies over Uruguay, also present in phases 4-5. The remaining pair of phases, 6-7 and 8-1, have only few events and may not represent correctly the impact of MJO. 
For events that start in phases 6-7 OLR anomalies over South America are strong only during phases 8-1: in extreme MJO events the SACZ strengthens, while during weak events northern South America shows stronger convection. 

For these events, there is a tendency that the SAD develops  only when the center of enhanced convection is located over the MC (phases 4-5) as reported by other authors \citep{diaz2020, grimm2019}.
 
Our results suggest the importance of the size and initiation phase of the MJO in inducing climate anomalies over South America during summer. As shown above, comparison of the evolution of extreme MJO events starting in phases 2-3 and 6-7 shows that OLR anomalies develop over some regions of South America irrespective of the initiation phase, while in other regions this is not the case. This suggests larger subseasonal predictability in the former regions. 

As for the impact of ENSO on the MJO we find that the duration of extreme MJO events during NT and LN years are longer than those during EN years. Moreover, MJO and ENSO can interact destructively or constructively depending on the phase of the former, decreasing the impact over South America. For example, extreme MJO events that start in phases 2-3 and that occur during EN present a clear SAD with decreased rainfall in the SACZ and increased rainfall over SESA from phases 2-3 up to phases 6-7. During NT years the  rainfall dipole can only be seen during phases 4-5. Also, extreme MJO events that start in phases 6-7 have a much stronger impact in the SACZ region when they occur during EN. During LN or NT years OLR anomalies over South America in phases 8-1 are much weaker. 

Overall, these results highlight the importance for monitoring the MJO, and particularly its initiation location and strength to help prediction of rainfall anomalies over South America on subseasonal time scales. Moreover, our findings  suggest the existence of predictions of opportunity based on the co-occurrence of the  ENSO and the MJO, and stress the need for further understanding the interaction between these two phenomena for skillful rainfall predictions. 

\vspace{0.9cm}

\textit{\textbf{Acknowledgments.}}
This work is part of the EU International Training Network (ITN) Climate Advanced Forecasting of subseasonal Extremes (CAFE) which has received funding from the European Union’s Horizon 2020 research and innovation programme under the Marie Skłodowska-Curie Grant Agreement No 813844.

\textit{\textbf{Data availability statement.}} The Real-Time Multivariate MJO index (RMM) was obtained from the Australian Bureau of Meteorology at \url{http://www.bom.gov.au/climate/mjo/}. The NOAA Interpolated Outgoing Longwave Radiation (OLR) data was provided by the NOAA PSL, Boulder, Colorado, USA, from their website at \url{https://psl.noaa.gov}.The ERA5 data used in this study is publicly available from the Copernicus Climate Data Store at \url{https://doi.org/10.24381/cds.f17050d7} as described in Hersbach et al. (2019). The ONI index is openly accessible and can be found on the website of the Physical Sciences Laboratory of NOAA at \url{https://psl.noaa.gov/data/climateindices/list/}.

\bibliographystyle{plainnat}
\bibliography{References}

\end{document}